\documentclass[twocolumn,amsmath,amssymb,showpacs,superscriptaddress,pra]{revtex4}

\usepackage{graphicx}
\usepackage{dcolumn}
\usepackage{bm}
\usepackage{graphicx,color}

\newcommand{\bs}{\boldsymbol}
\newcommand{\mb}{\mathbf}

\begin{document}

\title{Hidden momentum and the Abraham-Minkowski debate}

\author{Pablo L. Saldanha}\email{saldanha@fisica.ufmg.br}
\affiliation{Departamento de F\'isica, Universidade Federal de Minas Gerais, 30161-970. Belo Horizonte-MG, Brazil}
\author{J. S. Oliveira Filho}\email{juvenil@ufrb.edu.br}
\affiliation{Departamento de F\'isica, Universidade Federal de Minas Gerais, 30161-970. Belo Horizonte-MG, Brazil}
\affiliation{Universidade Federal do Rec\^oncavo da Bahia, 45300-000. Amargosa-BA, Brazil}


\begin{abstract}
We use an extended version of electrodynamics, which admits the existence of magnetic charges and currents, to discuss how different models for electric and magnetic dipoles do or do not carry hidden momentum under the influence of external electromagnetic fields. Based on that, we discuss how the models adopted for the electric and magnetic dipoles from the particles that compose a material medium influence the expression for the electromagnetic part of the light momentum in the medium. We show that Abraham expression is compatible with electric dipoles formed by electric charges and magnetic dipoles formed by magnetic charges, while Minkowski  expression is compatible with electric dipoles formed by magnetic currents and magnetic dipoles formed by electric currents. The expression $\varepsilon_0\mb{E}\times\mb{B}$, on the other hand, is shown to be compatible with electric dipoles formed by electric charges and magnetic dipoles formed by electric currents, which are much more natural models. So this expression has an interesting interpretation in the Abraham-Minkowski debate about the momentum of light in a medium: It is the expression compatible with the nonexistence of magnetic charges. We also provide a simple justification of why Abraham and Minkowski momenta can be associated with the kinetic and canonical momentum of light, respectively.  
\end{abstract}


\pacs{42.25.-p, 03.50.De}


\maketitle

\section{Introduction}

The expression for the momentum of light in a material medium has been discussed for more than 100 years  \cite{pfeifer07,barnett10b,milonni10,kemp11}. Abraham  expression $\mb{E}\times\mb{H}/c^2$ and Minkowski's   $\mb{D}\times\mb{B}$ for the electromagnetic momentum density were proposed in the beginning of the twentieth century. In these expressions, $\mb{E}$ is the electric field, $\mb{B}$ is the magnetic field, $\mb{D}=\varepsilon_0\mb{E}+\mb{P}$, $\mb{H}=\mb{B}/\mu_0 - \mb{M}$, $\mb{P}$ and $\mb{M}$ are the polarization and magnetization of the medium, $\varepsilon_0$ and $\mu_0$ are the permittivity and permeability of free space, and $c$ is the speed of light in vacuum. In a linear non-dispersive medium, the Minkowski expression predicts a light momentum proportional to the medium refractive index $n$, while Abraham's predicts a momentum inversely proportional to $n$. For a long time there was a discussion about which of these expressions was the correct one, but in the 1960's it was understood that a complete treatment for the problem must also include a material energy-momentum tensor, not only an electromagnetic energy-momentum tensor as in the original works from Abraham and Minkowski \cite{penfieldbook}. When the appropriate material tensor is taken into account both treatments, as well as others, predict the same experimental results  \cite{penfieldbook,pfeifer07}. The different treatments thus correspond to different divisions of the total energy-momentum tensor of the system into electromagnetic and material ones. 

Nowadays theoretical works on this subject investigate the physical meaning of the different expressions for the electromagnetic momentum \cite{barnett10a,saldanha10,saldanha11,leonhardt14}, as well as the experimental situations in which each one should manifest \cite{kemp11,pfeifer09,kemp11b,mansuripur12b,brevik12}. Modern experiments show how different electromagnetic momenta are relevant in different experimental situations \cite{campbell05,astrath14,zhang15,verma15}. In a recent work, Barnett discusses how Abraham  momentum is connected to the kinetic momentum of light and Minkowski  momentum is connected to the canonical momentum of light, and that this association is valid even in dispersive media \cite{barnett10a}. Besides Abraham and Minkowski momenta, we defend that there is another expression for the electromagnetic momentum density that has an interesting physical interpretation: $\varepsilon_0\mb{E}\times\mb{B}$. If the Lorentz force law is used to compute the momentum transfer from the electromagnetic wave to the material medium, this is the expression for the electromagnetic part of the momentum density which is compatible with momentum conservation \cite{saldanha10,saldanha11}. So this is the electromagnetic momentum density which is compatible with the Lorentz force law.

The difference between Abraham  expression $\mb{E}\times\mb{H}/c^2$ and $\varepsilon_0\mb{E}\times\mb{B}$ is $\mb{M}\times\mb{E}/c^2$, which can be associated with the hidden momentum density of the medium. Hidden momentum is a relativistic effect that may lead a magnetic dipole $\bs{\mu}$ in the presence of an external electric field $\mb{E}$ to acquire a linear momentum $\bs{\mu}\times\mb{E}/c^2$ even if the dipole is not moving \cite{babson09}. The hidden momentum concept is intimately connected to the Lorentz force law. For instance, the Mansuripur paradox \cite{mansuripur12}, which questions the validity of the Lorentz force law due to the fact that it predicts a nonzero torque from an electric field in a magnetic dipole in one reference frame, but a nonzero torque in other frames, is solved when the hidden momentum of the dipole is taken into account \cite{saldanha13,vanzella13,khorrami13,barnett13}. Until recently there were only classical models for hidden momentum \cite{penfieldbook,shockley67,vaidman90,hnizdo97,babson09}, but we have used perturbation theory to show that a hydrogen atom with magnetic moment due to the electron orbital angular momentum carries hidden momentum in the presence of an external electric field \cite{juvenil15}. So, since atoms carry hidden momentum and matter is made of atoms, the magnetic dipoles induced in the medium by the presence of an electromagnetic wave should carry hidden momentum under the influence of the wave electric field, resulting in a hidden momentum density given by $\mb{M}\times\mb{E}/c^2$.

In this work we discuss that the difference between the kinetic electromagnetic momentum density $\mb{E}\times\mb{H}/c^2$ and the expression $\varepsilon_0\mb{E}\times\mb{B}$ results from the consideration of the hidden momentum density $\mb{M}\times\mb{E}/c^2$ as being part the electromagnetic momentum density or as being part of the material momentum density. We generalize Maxwell equations and the Lorentz force law considering the hypothetical existence of magnetic charges and currents. In this way, an electric dipole formed by magnetic currents should also carry hidden momentum in the presence of a magnetic field. We then discuss which expressions for the electromagnetic momentum density are compatible with different models for the electric and magnetic dipoles of the medium. Our discussion is based on the fact that the existence or absence of hidden momentum in the particles depends on the adopted models. Since the total momentum must be the same independently of the model, the electromagnetic part of the momentum also depends on the model. In particular, we show that Abraham  expression is compatible with electric dipoles formed by electric charges and magnetic dipoles formed by magnetic charges. We also show that Minkowski  expression is compatible with electric dipoles formed by magnetic currents and magnetic dipoles formed by electric currents. The expression $\varepsilon_0\mb{E}\times\mb{B}$, on the other hand, is shown to be compatible with electric dipoles formed by electric charges and magnetic dipoles formed by electric currents, which are much more natural models. So this expression for the electromagnetic momentum density certainly has its value in the Abraham-Minkowski debate.

This paper is organized as follows. In Sec. \ref{sec:am} we present a simple explanation of why Abraham and Minkowski momenta can be associated with the kinetic and canonical electromagnetic momenta, respectively. In Sec. \ref{sec:hm} we generalize Maxwell equations and the Lorentz force law considering the existence of magnetic charges and currents, and discuss which models for electric and magnetic dipoles carry or do not carry hidden momentum in the presence of external electromagnetic fields. In Sec. \ref{sec:hmam} we discuss how different models for the electric and magnetic dipoles responsible for the polarization and magnetization of a material medium are connected with different expressions for the electromagnetic momentum density. Finally, in Sec. \ref{sec:conc} we present our concluding remarks. The treatments of Secs. \ref{sec:am} and \ref{sec:hmam} are the original contributions of the present work.

\section{Physical interpretation of Abraham  and Minkowski  momenta} \label{sec:am}

Consider a neutral particle with mass $m$, magnetic dipole moment $\bs{\mu}_0$ and electric dipole moment $\mb{p}_0$ in its rest frame. The non-relativistic Lagrangian that governs the interaction of this particle with external electromagnetic fields is 
\begin{equation}
	\mathcal{L}=\frac{1}{2}mv^2+\mb{E}\cdot\mb{p}+\mb{B}\cdot\bs{\mu},
\end{equation}
where $\mb{p}=\mb{p}_0+\mb{v}\times\bs{\mu}_0/c^2$ and $\bs{\mu}=\bs{\mu}_0-\mb{v}\times\mb{p}_0$ are the electric and magnetic dipole moments when the particle has velocity $\mb{v}$ \cite{panofsky,hnizdo12}. Each component of the particle canonical momentum is given by
\begin{equation}
	\frac{\partial\mathcal{L}}{\partial{v_i}}=mv_i-(\mb{p}_0\times\mb{B})_i+\frac{(\bs{\mu}_0\times\mb{E})_i}{c^2},
\end{equation}
while its (non-relativistic) kinetic momentum is $m\mb{v}$.

If we consider a medium composed of several neutral particles, after taking averages over a volume that contains many of these particles, according to the above treatment the difference between the canonical momentum density $\mb{P}_\mathrm{can}$ and the kinetic momentum density $\mb{P}_\mathrm{kin}$ of the medium is 
\begin{equation}
	\mb{P}_\mathrm{can}-\mb{P}_\mathrm{kin}=-\mb{P}\times\mb{B}+\frac{\mb{M}\times\mb{E}}{c^2},
\end{equation}
where $\mb{P}$ and $\mb{M}$ represent the polarization and magnetization of the medium, respectively. We are assuming that the average velocity of the particles is small, such that the contribution for the medium magnetization and polarization comes from the electric and magnetic dipoles from each particle in its rest frame. In a gaseous medium, the contribution of each particle velocity for its own magnetic and dipole moments may be non-negligible, but since the particles velocities are random, the average contribution from the velocities of all particles on the medium magnetization and polarization is assumed to be negligible. 

The difference between Minkowski momentum density $\mb{P}_\mathrm{Min}=\mb{D}\times\mb{B}$ and Abraham  momentum density $\mb{P}_\mathrm{Abr}=\mb{E}\times\mb{H}/c^2$ is 
\begin{equation}
	\mb{P}_\mathrm{Min}-\mb{P}_\mathrm{Abr}=\mb{P}\times\mb{B}-\frac{\mb{M}\times\mb{E}}{c^2},
\end{equation}
such that we can write
\begin{equation}
	\mb{P}_\mathrm{can}+\mb{P}_\mathrm{Min}=\mb{P}_\mathrm{kin}+\mb{P}_\mathrm{Abr}.
\end{equation}
We can interpret the above formula as saying that the total momentum of an electromagnetic wave in a medium can be written as the sum of the canonical material momentum and Minkowski  momentum or as the sum of the kinetic material momentum and Abraham  momentum. This reflects the fact that the system total momentum can be divided into electromagnetic and material parts in different ways. So Minkowski  momentum can be associated with the canonical electromagnetic momentum and Abraham  momentum with the kinetic electromagnetic momentum.

This association of the Abraham and Minkowski momenta as the kinetic and canonical electromagnetic momenta, respectively, is supported by experiments and other theoretical considerations. Experiments that measured the radiation pressure on atoms from a Bose-Einstein condensate showed that the radiation pressure is proportional to the medium refractive index $n$, being compatible with Minkowski  momentum \cite{campbell05}. This is expected, since quantum systems must be treated with a Hamiltonian formalism where the canonical momentum plays the crucial role. A treatment from first principles for the photon radiation pressure on a mirror whose position is treated quantum mechanically also showed that the radiation pressure is proportional to $n$ \cite{correa16}.  Experiments that measured the radiation pressure of classical light in classical mirrors immersed in different media indeed showed that the radiation pressure is proportional to $n$ \cite{jones51,jones78}. On the other hand, in the discussion of the movement of a transparent block due to the transmission of an electromagnetic pulse, considering the uniformity of the movement of the system center of energy, it is the Abraham momentum that appears, since the phenomenon is kinetic \cite{balazs53,loudon04,ramos11}. 

In this section we have used a non-relativistic Lagrangian to arrive at our conclusions. However, in the following sections a relativistic treatment must be used to include hidden momentum, which is a relativistic effect. The overall idea of our treatment is that the movement of the particles that compose the atoms must be treated relativistically, as in our previous work that describes hidden momentum in a hydrogen atom \cite{juvenil15}, but we consider that the atoms that compose the medium have non-relativistic velocities such that their movement can be treated classically, as in this section. 

\section{Hidden momentum of electric and magnetic dipoles}\label{sec:hm}

Different models for magnetic dipoles carry hidden momentum under the influence of an external electric field \cite{penfieldbook,shockley67,vaidman90,hnizdo97,babson09,juvenil15}. These models have in common the fact that the external field does work on the circulating electric charges that compose the magnetic dipole. This work is positive in some parts of the circuit and negative in others, changing the charges energies along the circuit. In a relativistic description, the difference of the moving charges energies in different portions of the circuit generates a linear momentum given by $\bs{\mu}\times\mb{E}/c^2$ even if the dipole is not moving \cite{babson09}. Since an electric dipole is formed by a charge distribution in which the center of the positive charges is dislocated in relation to the center of the negative charges, it does not carry hidden momentum under the influence of external electromagnetic fields, as the fields do no work on the electric charges that compose the dipole in its rest frame.  

The lack of experimental evidence attesting to the reality of magnetic charges does not prevent us from assuming the existence of these in a purely theoretical way. Actually, the extension of the electromagnetic theory including these entities is straightforward. Considering the existence of magnetic monopoles, the Maxwell equations for a physical system consisting of electric and magnetic charges and currents can be written as \cite{griffithsbook}
\begin{equation}
\mb{\nabla} \cdot \mb{E} = \frac{\rho_{e}}{\epsilon_{0}}, \label{eq1} 
\end{equation}
\begin{equation}
\mb{\nabla} \cdot \mb{B} = \mu_{0} \rho_{m}, \label{eq2} 
\end{equation}
\begin{equation}
\mb{\nabla} \times \mb{E} = - \mu_{0} \mb{J}_{m} -\frac{\partial \mb{B}}{\partial t}, \label{eq3} 
\end{equation}
\begin{equation}
\mb{\nabla} \times \mb{B} = \mu_{0} \mb{J}_{e} + \mu_{0} \epsilon_{0} \frac{\partial \mb{E}}{\partial t}, \label{eq4} 
\end{equation}
where $\rho_{e}$  and $\rho_{m}$ represent the electric and magnetic charge densities, and  $\mb{J}_{e}$ and $\mb{J}_{m}$ the electric and magnetic current densities, respectively. Moreover, the general expression for the force density, which describes how the electric and magnetic fields exert force on the charges and currents of the system, may be written as
\begin{equation}
 \mb{f} = \rho_{e} \mb{E} + \mb{J}_e \times \mb{B} + \rho_{m} \mb{B} - \frac{1}{c^{2}} \mb{J}_{{m}} \times \mb{E} . \label{eq5} 
\end{equation}

If we consider a magnetic dipole formed by separated magnetic charges of opposite sign, this system will not carry hidden momentum under the influence of an external electric field, since there will be no moving parts for the external field to do work on. On the other hand, an electric dipole $\mb{p}_{0}$ formed by a circulating magnetic current will carry hidden momentum in the presence of an external magnetic field $\mb{B}$, since the magnetic field will do work on the circulating magnetic current. The symmetry of Maxwell equations and the force law with electric and magnetic charges from Eqs. (\ref{eq1})-(\ref{eq5}) tells us that the hidden momentum will be $ -\mb{p}_{0} \times \mb{B}$ in this case. So the existence or absence of hidden momentum in electric and magnetic dipoles under the influence of external electromagnetic fields depend on the models we adopt for the dipoles.

\section{Connection between the medium model and the electromagnetic momentum density}\label{sec:hmam}

\begin{table*}
\caption{\label{tab:table1}Association of different models for the electric dipoles $\mb{p}_i$ and magnetic dipoles $\boldsymbol{\mu}_i$ of the particles that compose the medium with the corresponding expressions for the electromagnetic momentum density.}
\begin{ruledtabular}
\begin{tabular}{cccc}
Model for $\mb{p}_i$ & Model for $\boldsymbol{\mu}_i$ & Hidden momentum density & Electromagnetic momentum density \\
\hline
Electric charges & Magnetic charges & 0 & $\mb{E}\times\mb{H}/c^2$ \\
Electric charges & Electric currents & $\mb{M}\times\mb{E}/c^2$ & $\varepsilon_0\mb{E}\times\mb{B}$ \\
Magnetic currents & Electric currents & $-\mb{P} \times \mb{B}+\mb{M}\times\mb{E}/c^2$ & $\mb{D}\times\mb{B}$ \\
Magnetic currents & Magnetic charges & $-\mb{P} \times \mb{B}$ & $\mb{E}\times\mb{H}/c^2+\mb{P}\times\mb{B}$ \\\end{tabular}
\end{ruledtabular}
\end{table*}

In this section we discuss how different models for the electric and magnetic dipoles responsible for the polarization and magnetization of a material medium are connected with different expressions for the electromagnetic part of the momentum of light in the medium.

First let us consider that a medium is composed by neutral particles whose electric dipole moments are the result of the separation of electric charges and the magnetic dipole moments are the result of the separation of magnetic charges. According to the discussion of Sec. \ref{sec:hm}, these particles do not carry hidden momentum in the presence of electromagnetic fields, such that their momentum is purely kinetic. So this model is compatible with Abraham momentum for the electromagnetic field, which is the electromagnetic kinetic momentum. The system total momentum is thus divided into kinetic material momentum and kinetic electromagnetic momentum.

Considering now that a medium is composed of neutral particles whose electric dipole moments $\mb{p}_i$ are the result of the separation of electric charges and the magnetic dipole moments $\bs{\mu}_i$ are the result of electric currents, according to the discussion of Sec. \ref{sec:hm} each particle carries a hidden momentum $\bs{\mu}_i\times\mb{E}/c^2$ in the presence of an electric field $\mb{E}$. So the material momentum is composed of the particles kinetic momentum plus the particles hidden momentum. Since the total momentum must be the same as before, we must subtract the hidden momentum from the Abraham electromagnetic momentum to obtain the electromagnetic momentum compatible with these models. The electromagnetic momentum density in this case becomes $\mb{E}\times\mb{H}/c^2-\mb{M}\times\mb{E}/c^2=\varepsilon_0\mb{E}\times\mb{B}$. The difference between these expressions thus comes from the consideration of $\mb{M}\times\mb{E}/c^2$ as being part of the electromagnetic momentum density (in $\mb{E}\times\mb{H}/c^2$) or as being part of the material momentum density (in $\varepsilon_0\mb{E}\times\mb{B}$). Since the usual expression of the Lorentz force law (without magnetic charges or currents) is connected to the existence of hidden momentum in magnetic dipoles under the influence of electric fields, we see why $\varepsilon_0\mb{E}\times\mb{B}$ is the expression for the electromagnetic momentum density which is compatible with the use of the Lorentz force law to compute the momentum transfer from the wave electromagnetic fields to the material medium \cite{saldanha10,saldanha11}. 

For a medium composed of neutral particles whose electric dipole moments $\mb{p}_i$ are the result of magnetic currents and the magnetic dipole moments $\bs{\mu}_i$ are the result of electric currents, each particle carries a hidden momentum $\bs{\mu}_i\times\mb{E}/c^2 -\mb{p}_i \times \mb{B}$ in the presence of electromagnetic fields $\mb{E}$ and $\mb{B}$. Again, the material momentum is composed of the particles kinetic momentum plus the particles hidden momentum. Following the same argument as before, we must subtract the hidden momentum from the Abraham electromagnetic momentum to obtain the electromagnetic momentum compatible with these models. The electromagnetic momentum density in this case becomes $\mb{E}\times\mb{H}/c^2-\mb{M}\times\mb{E}/c^2+\mb{P}\times\mb{B}=\mb{D}\times\mb{B}$, which is the Minkowski momentum density.

In our last model, consider  a medium composed of neutral particles whose electric dipole moments $\mb{p}_i$ are the result of magnetic currents and the magnetic dipole moments $\bs{\mu}_i$ are the result of the separation of magnetic charges. In this case, each particle carries a hidden momentum $-\mb{p}_i \times \mb{B}$ in the presence of a magnetic field $\mb{B}$. Subtracting this hidden momentum from the Abraham electromagnetic momentum, the electromagnetic momentum density in this case becomes $\mb{E}\times\mb{H}/c^2+\mb{P}\times\mb{B}$.

The results of this section are summarized in Table I.

\section{Discussion} \label{sec:conc}

As we have shown in Sec. \ref{sec:hmam}, Abraham momentum is compatible with a model for the material medium in which the electric dipoles are formed by electric charges and the magnetic dipoles are formed by magnetic charges. Minkowski momentum is compatible with a model for the material medium in which the electric dipoles are formed by magnetic currents and the magnetic dipoles are formed by electric currents. The expression $\varepsilon_0\mb{E}\times\mb{B}$ for the electromagnetic momentum density, on the other hand, is compatible with a model for the material medium in which the electric dipoles are formed by electric charges and the magnetic dipoles are formed by electric currents, which is a much more natural model. So this expression also has an interesting interpretation as an electromagnetic momentum density, besides Abraham momentum as the kinetic momentum and Minkowski momentum as the canonical momentum. It is the expression compatible with the Lorentz force law and with the nonexistence of magnetic charges.   

Since it was recently shown that a hydrogen atom with magnetic moment due to the orbital angular momentum of the electron carries hidden momentum in the presence of an electric field \cite{juvenil15}, the expression $\varepsilon_0\mb{E}\times\mb{B}$ is also compatible with a quantum atomic model for the material medium. Despite the fact that, to our knowledge, there is no treatment that shows the existence or absence of hidden momentum for the electron spin in the presence of an electric field, the magnetic responses of matter due to the electron spin are slow, resulting from relaxation processes. For optical fields with high frequencies, the magnetic responses of transparent media are related to the orbital angular momentum of the electrons \cite{vanvleck,saldanha11}, such that in these situations the assumption of the existence of hidden momentum in the medium is valid.

We acknowledge Daniel Vanzella and Stephen Barnett for very useful discussions. This work was supported by the Brazilian agencies CNPq and CAPES.


\begin{thebibliography}{}



\bibitem{pfeifer07} 
R. N. C. Pfeifer, T. A. Nieminen, N. R. Heckenberg, and H. Rubinsztein-Dunlop,
Rev.\ Mod.\ Phys.\ {\bf 79}, 1197 (2007).
  
\bibitem{barnett10b}
S. M. Barnett and R. Loudon,
Phil.\ Trans.\ R.\ Soc.\ A\ {\bf 368}, 927 (2010).

\bibitem{milonni10}
P. W. Milonni and R. W. Boyd, Adv. Opt. Phot. \textbf{2}, 519 (2010).	


\bibitem{kemp11}
B. A. Kemp, J. Appl. Phys. \textbf{109}, 111101 (2011).

	
\bibitem{penfieldbook}
P. Penfield Jr. and H. A. Haus,
\textit{Electrodynamics of Moving Media}
(MIT, Cambridge, 1967).  
 
\bibitem{barnett10a} 
S. M. Barnett,
Phys.\ Rev.\ Lett.\  {\bf 104}, 070401 (2010).

\bibitem{saldanha10}
P. L. Saldanha, 
Opt.\ Express\  {\bf 18}, 2258 (2010).

\bibitem{saldanha11}
P. L. Saldanha, 
Opt.\ Commun.\  {\bf 284}, 2653 (2011).

\bibitem{leonhardt14}
U. Leonhardt, Phys. Rev. A \textbf{90}, 033801 (2014).


\bibitem{pfeifer09}
R. N. C. Pfeifer, T. A. Nieminen, N. R. Heckenberg, and H. Rubinsztein-
Dunlop, Phys. Rev. A \textbf{79}, 023813 (2009).


\bibitem{kemp11b}
B. A. Kemp and T. M. Grzegorczyk, Opt. Lett. \textbf{36}, 493 (2011).

\bibitem{mansuripur12b}
M. Mansuripur, Phys. Rev. A \textbf{85}, 023807 (2012).

\bibitem{brevik12}
I. Brevik and S. A. Ellingsen, Phys. Rev. A \textbf{86}, 025801 (2012).


\bibitem{campbell05}
G. K. Campbell \textit{et al.}, Phys. Rev. Lett. \textbf{94} 170403 (2005).

\bibitem{astrath14}
N. G. C. Astrath \textit{et al.}, Nat. Comms. \textbf{5}, 4363 (2014).

\bibitem{zhang15}
L. Zhang, W. She, N. Peng, and U. Leonhardt, New J. Phys. \textbf{17}, 053035 (2015).

\bibitem{verma15}
G. Verma and K. P. Singh, Phys. Rev. Lett. \textbf{115}, 143902 (2015).

\bibitem{babson09}
D. Babson, S. P. Reynolds, R. Bjorkquist, and D. J. Griffiths, 
Am.\ J.\ Phys.\  {\bf 77}, 826 (2009).


\bibitem{mansuripur12}
M. Mansuripur, 
Phys.\ Rev.\ Lett.\  {\bf 108}, 193901 (2012).


\bibitem{vanzella13}
D. A. T. Vanzella, 
Phys.\ Rev.\ Lett.\  {\bf 110}, 089401 (2013).

\bibitem{barnett13}
S. M. Barnett, 
Phys.\ Rev.\ Lett.\  {\bf 110}, 089402 (2013).

\bibitem{saldanha13}
P. L. Saldanha, 
Phys.\ Rev.\ Lett.\  {\bf 110}, 089403 (2013).

  
\bibitem{khorrami13}
M. Khorrami, 
Phys.\ Rev.\ Lett.\  {\bf 110}, 089404 (2013).



\bibitem{shockley67}
W. Shockley and R. P. James, 
Phys.\ Rev.\ Lett.\  {\bf 18}, 876 (1967).  

  
\bibitem{vaidman90}
L. Vaidman, 
Am.\ J.\ Phys.\  {\bf 58}, 978 (1990).	
	
\bibitem{hnizdo97}
V. Hnizdo, 
Am.\ J.\ Phys.\  {\bf 65}, 55 (1997).


\bibitem{juvenil15}
J. S. Oliveira Filho and P. L. Saldanha, Phys Rev. A \textbf{92}, 052107 (2015).

\bibitem{panofsky}
K. W. H. Panofsky and M. Phillips, \textit{Classical Electricity and Magnetism} (Dover, New York, 2005), 2nd ed. Sec. 18-4.

\bibitem{hnizdo12}
V. Hnizdo, Am. J. Phys. \textbf{80}, 645 (2012).

\bibitem{correa16}
R. Corr\^ea and P. L. Saldanha, Phys. Rev. A \textbf{93}, 023803 (2016).


\bibitem{jones51}
R. V. Jones, Nature \textbf{167}, 439 (1951).

\bibitem{jones78}
R. V. Jones and B. Leslie, Proc. R. Soc. Lond. A \textbf{360}, 347 (1978).


\bibitem{balazs53} 
N. L. Balazs, Phys. Rev. \textbf{91}, 408 (1953). 

\bibitem{loudon04}
R. Loudon, Fortschr. Phys. \textbf{52}, 1134 (2004).

\bibitem{ramos11}
T. Ramos, G. F. Rubilar and Y. N. Ubukhov, Phys. Lett. A \textbf{375}, 1703 (2011).


\bibitem{griffithsbook}
D. J. Griffiths,
\textit{Introduction to Electrodynamics} (Prentice Hall, Upper Saddle River, 1999), 3rd ed.


 \bibitem{vanvleck}
J. H. van Vleck,
\textit{The Theory of Electric and Magnetic Susceptibilities} (Oxford University Press, London, 1966).

  
\end{thebibliography}
\end{document}